\def\vec{}
\def\gradient{\nabla}

\def\ang{l}
\def\<<{{\ll}}
\def\>>{{\gg}}
\def\spose#1{\hbox to 0pt{#1\hss}}
\def\ltwig{\mathrel{\spose{\lower 3pt\hbox{$\mathchar"218$}}
     \raise 2.0pt\hbox{$\mathchar"13C$}}}
\def\gtwig{\mathrel{\spose{\lower 3pt\hbox{$\mathchar"218$}}
     \raise 2.0pt\hbox{$\mathchar"13E$}}}
\def\blankline{\par\vskip \baselineskip}
\def\beq{\begin{equation}}
\def\eeq{\end{equation}}

\def\>>{\gg}

\def\=={\equiv}

\def\xfun{X}
\def\yfun{Y}

\documentstyle[12pt,aaspp4,flushrt]{article}

\begin{document}


\title{Line Forces in Keplerian Circumstellar Disks and Precession of Nearly Circular Orbits}

\author{K. G. Gayley}
\affil{Department of Physics and Astronomy, University of Iowa, Iowa City, IA 52242}

\author{R. Ignace}
\affil{Department of Physics and Astronomy, University of Iowa, Iowa City, IA 52242}

\author{S. P. Owocki}
\affil{Bartol Research Institute, University of Delaware, Newark, DE 19716}

\begin{abstract}

We examine the effects of optically thick line forces on orbiting
circumstellar disks, such as occur around Be stars.  For radially streaming
radiation (e.g., as from a point source), line forces are only effective if
there is a strong radial velocity gradient, as occurs, for example, in a
line-driven stellar wind.  However, we emphasize here that, within an
orbiting disk, the radial shear of the azimuthal velocity leads to strong
line-of-sight velocity gradients along nonradial directions.  As
such, we show that, in the proximity of a stellar surface extending over a
substantial cone angle, the nonradial components of stellar radiation can
impart a significant line force to such a disk, even in the case of purely
circular orbits with {\it no} radial velocity.

Given the highly supersonic nature of orbital velocity variations, we use
the Sobolev approximation for the line transfer, extending to the disk case
the standard CAK formalism developed for line-driven winds.  We delineate
the parameter regimes for which radiative forces might alter disk
properties; but even when radiative forces are small, we analytically quantify
higher-order effects in the linear limit, including the
precession of weakly elliptical orbits.  We find that optically thick line
forces, both radial and azimuthal, can have observable implications for the
dynamics of disks around Be stars, including the generation of either
prograde or retrograde precession in slightly eccentric orbits.  However,
our analysis here suggests a net {\it retrograde} effect, in apparent
contradiction with observed long-term variations of violet/red line profile
asymmetries from Be stars, which are generally thought to result from {\it
prograde} propagation of a one-arm, disk-oscillation mode.  We also
conclude that radiative forces may alter the dynamical properties at the
surface of the disk where disk winds originate, and in the outer regions
far from the star, and may even make low-density disks vulnerable to
being blown off completely.

\end{abstract}


\section{Introduction}


Be stars
are defined to be non-supergiant B stars that have shown
emission in one or more Balmer lines at some time.  As early as 1931,
Struve had identified non-sphericity associated with rapid stellar
rotation as the source for the Be star emission lines and profile
shapes.  Although the details have changed substantially, Struve's
overall premise has withstood the test of time.  Optical interferometry
has conclusively shown that Be stars possess extended equatorial
circumstellar disks that are geometrically thin, with the opening angle
of the inner disk being of order a few degrees to be consistent with
interferometric and 
polarimetric measurements (Quirrenbach et al. 1997; Wood,
Bjorkman, \& Bjorkman 1997).

Increasing observational evidence suggests that these
circumstellar disks are in 
Keplerian orbit (e.g., Waters \& Marlborough 1994; Hummel \& Vrancken 2000).
%
For example, Keplerian disks have been shown to
be capable of supporting one-armed spiral density modes, whereas an
angular-momentum conserving disk will not (Okazaki 1991, 1996, 1997).
These one-armed modes 
result in a non-axisymmetric disk structure that is supported by
observations (Telting et al. 1994; Hanuschik et al. 1995; Hummel \& Hanuschik 1997).

The one-armed pattern is also typically observed to precess around the disk.
It has long been known that some Be stars that
show violet (V) and red (R) peaks with a central depression 
in the H$\alpha$ emission line can
undergo quasi-periodic variations in the relative strengths of these peaks
(e.g., Dachs 1987; Hubert 1994).  This cyclic
behavior is commonly referred to as ``V/R variations'' with typical
periods of years to decades (e.g., see the compilation in the Appendix
of Okazaki 1997).  The V/R variations are thought to result from
the prograde precession of ``Global Disk
Oscillations'' (GDOs) in Keplerian Be disks (Papaloizou, Savonije,
\& Henrichs 1992; Telting et al. 1994; Hanuschik 1994; 
Mennickent, Sterken, \& Vogt 1997).
Measurements with optical
interferometry have now confirmed the non-axisymmetric disk structure
in $\zeta$ Tau and $\gamma$ Cas, and have even observed the slow
secular motion of the pattern around the star (Vakili et al. 1998;
Berio et al. 1999).

However, there are numerous outstanding concerns in relation to Be disks
and GDOs.  It is unclear exactly how the
Keplerian disks are generated, and two well-studied Be stars
($\lambda$ Eri: Smith et al. 1993; and $\gamma$ Cas: Robinson \& Smith 2000;
Cranmer, Robinson, \& Smith 2000), show evidence for flaring activity and co-rotating
clouds of gaseous material that are interpreted as signatures of
magnetic fields.
Furthermore, several Be stars are
observed to be non-radial pulsators, and so variations in temperature and
brightness across the face of the star might affect the disk structure.
Finally, the GDO behavior is actually
only quasi-periodic, and
sometimes the V/R variations 
can change period, appear, or disappear entirely.
Even the entire circumstellar disk can ``disappear'', only to be
regenerated at a later time.  For example the disk in the Be star $\mu$
Cen appears to be ``rebuilding'' itself (Rivinius et al. 1998), at a rate
consistent with stellar mass loss (Telting et al.
1998).
The lack of a clear explanation for such curious
behavior suggests that the study of Be stars is in a phase where
all the relevant physics is still being assembled, and all potentially
important contributors must be considered.

In this spirit, we consider here
a process that has been generally overlooked:
the potential impact of
{\it optically thick line} forces on Keplerian disks and
GDOs.  In the past, GDO models have either
ignored radiative forces (e.g., Papaloizou, Savonije, \& Henrichs 1992) or assumed that
the radiative line driving force is purely radial and derives entirely
from optically thin lines (e.g., Okazaki 1997, which adopts the 
purely parametric prescription
of Chen \& Marlborough 1994).  However, 
since the line force is affected by the strong velocity shear present in
a Keplerian disk,
the highly supersonic orbital speed effectively broadens optically thick
lines. 
Much as in stellar winds, this can have important dynamical consequences.

To understand how a Keplerian disk generates line-of-sight velocity gradients,
consider the schematic in Figure 1.
The key point is that the finite size of the star allows for nonradial
radiation streams, including the tangential stream from the limb depicted
in the figure.
These streams encounter the disk at oblique angles that sample the Keplerian
velocity gradient, even in the absence of any radial motions.
Since the Keplerian speeds are of the same order of magnitude as wind speeds,
and they vary over the same scale, the line-of-sight gradients can also be
of the same order for favorable geometries.
Thus the geometric differences between applying CAK-type line forces 
(Castor, Abbott, \& Klein 1975) to
winds and disks are not as large as has been assumed in the past, and indeed
the primary difference will be shown to be related to the density contrast,
not the geometry.

\begin{figure}
\unitlength1in
\begin{picture}(6.5,5.)
   \leavevmode
   \epsfxsize=6.5in
   \epsffile{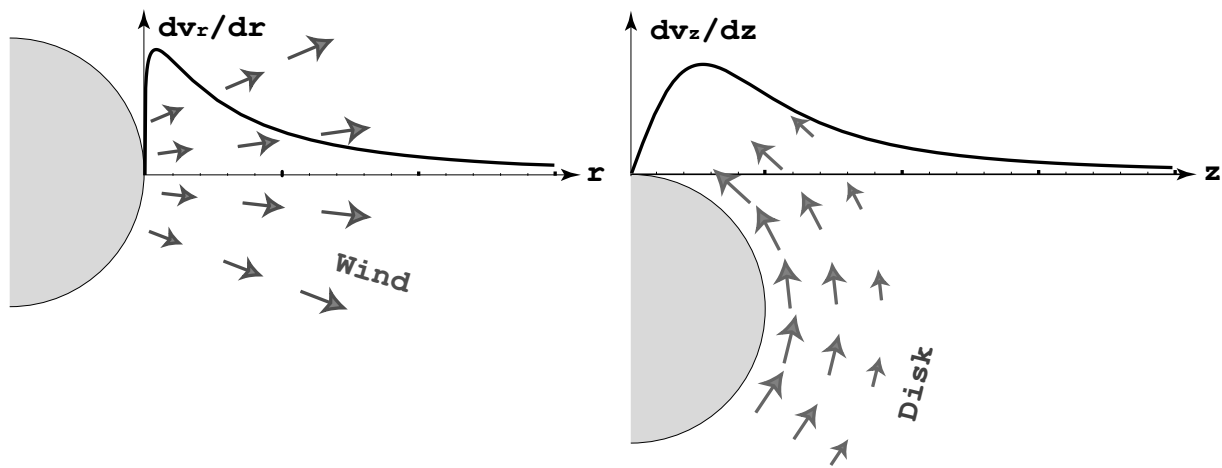}
\end{picture}
\hfill
\caption[ ]{\footnotesize
Comparison of the radial velocity gradient, $dv_r/dr$, of a radial stellar
wind outflow with the line-of-sight gradient, $dv_z/dz$, along
a ray direction $z$ that is {\it tangent}
to the stellar limb and passing through an orbiting
Keplerian disk.
The units of the gradient are arbitary, but the
relative scaling is accurate for the case in which the terminal velocity of
the wind is the same as the near-surface orbital velocity of the disk.
Note that, even without any radial outflow, the gradients are quite comparable.
This represents a key factor in the potential for line-forces to be important
in orbiting disks as well as outflowing winds.
\label{fig1}
}
\end{figure}

The magnitude of the line-driving force 
cannot dominate the overall gravitational binding if disks are to exist
in steady state.
However, it can readily compete with the relatively
weak forces commonly 
included in models of higher-order disk dynamics, 
in which GDO precession is
controlled by
pressure and nonspherical corrections to the
stellar gravity (Okazaki 1997; Mennickent, Sterken, \& Vogt 1997).
Previous models have aleady found that the form of the linearized GDO solution
can be altered by optically thin line-forces, thus leading to
different responses in
early- and late-type
Be stars (Okazaki 1997, 2000).  
Here we provide a substantially more general treatment of disk line forces,
built upon the formalism developed Castor, Abbott, \& Klein (1975)
to model line-driven stellar winds.

Although this paper only specifically treats cricumstellar disks around hot
stars, the formalism may be applicable to other UV-bright disk environments,
such as accretion disks in cataclysmic variables 
(Drew \& Proga 2000; Feldmeier, Shlosman, \& Vitello
1999).
Another potentially important analogy is with accretion disks thought to exist in
active galactic nuclei.
Recently, line forces in such disks have been considered by Proga, Stone,
\& Kallman (2000), and such work can also benefit from the analytic perspective
taken in this paper, tailored to that alternate context.

In \S 2, we describe our approach for extending the CAK line-force formalism to 
the case of a Keplerian disk, and then in \S 3 estimate the magnitude of the
resulting forces for a range of model parameters.
These estimates are often small compared to the radial force scale set by
gravity, and so in \S 4 we investigate in the linear limit the higher-order effect on 
low-amplitude disk perturbations, deriving the precession for a homologous family
of slightly eccentric orbits.
The conclusions of how line forces affect low-density regions of the disk and
the precession rates of nearly circular orbits are given in \S 5.
An Appendix gives the details of the analytic
calculation of the precession period.

\section{Treatment of Line Forces}

Within the context of the above introduction, let us now lay the basis
for quantifying line forces in Keplerian disks.
The essential physics of supersonic 
line driving involves the Doppler shifting of
lines into resonance with stellar continuum photons 
over a narrow
region of physical space.
This allows each such line to be treated in the Sobolev approximation
(Sobolev 1960), and the cumulative effect of the thousands of 
contributing lines can then be accomodated using the standard CAK formalism.
This formalism is by now quite well established, and so we will merely
cite the results relevant for the forces on gas parcels in nearly
circular orbits in the equatorial plane, taking the 
geometric expressions derived
in Cranmer \& Owocki (1995) and using the nomenclature
described in Gayley (1995).

To accomodate the effects of thousands of arbitrary lines, we follow a
statistical approach equivalent to CAK.
We fit the number of lines ${\tilde N}(q)$ with line strength greater
than $q$ with a power
law over the domain of marginally
thick lines, which is decisive for the line force.
The line strength $q$ is here given for line $i$ to be
\beq
q_i \ = \ {3 
\over 8} {\lambda_i \over r_e} a_i f_i \ ,
\eeq
where $\lambda_i$ is the line wavelength, $r_e = e^2/mc^2$ is the 
classical radius of the electron, $a_i$ is the abundance, relative
to free electrons, of bound
electrons in the lower level of the transition, and $f_i$ is the
oscillator strength.
The value of $q$ may be interpreted as the force on the bound electrons,
relative to the force on the free electrons, with no account taken
for opacity effects that redistribute photons out of the absorbing
bandwidth and limit the radiative acceleration.

To obtain the standard meaning of the CAK line-list parameter $\alpha$,
we assume
\beq
\label{tildenq}
{\tilde N}(q) \ = \  \left ( { q \over Q} \right )^{\alpha - 1} \ ,
\eeq
where $Q$ is simply a normalization constant that can be
interpreted as the
$q$ of the apparent strongest line, if the power-law distribution is
extrapolated from the domain of marginally thick lines into the realm
of extremely thick lines.
Note $Q$ is not necessarily the $q$ of the {\it actual} strongest line, which
plays little role in the total radiative force.
For readers familiar with prior literature
in this area, note that it is more standard to define a
{\it differential} line-strength distribution function, of which
${\tilde N}(q)$, having the meaning of an actual number
of lines, is the integrated form.

This simple form for ${\tilde N}(q)$ yields for the radiative acceleration
in vector notation (see e.g.,
Cranmer \& Owocki 1996; Owocki, Cranmer, \& Gayley 1996)
\beq
\label{gradgeneral}
{\bf {\vec g}} \ = \ \Gamma(\alpha) Q {\kappa_{es} \over c} 
\int d{\hat n} \ {\bf {\hat n}} I({\bf {\hat n}}) 
\left [ {\tilde t}({\bf {\hat n}}) \right ]^{-\alpha} \ , 
\eeq
where we have introduced 
the line-shadowing parameter 
\beq
\label{tildet}
{\tilde t}({\bf {\hat n}}) \ = \  Q \rho \kappa_{es} c
\left [ \left ( {\bf {\hat n}} \cdot \gradient \right )
\left ( {\bf {\hat n}} \cdot {\bf {\vec v}} \right ) \right ]^{-1} \ ,
\eeq
where $\Gamma(\alpha)$ is the standard Gamma function.
Note that ${\tilde t}$
can be thought of as the Sobolev optical depth of a line of strength
$Q$.
For a power-law line list, ${\tilde t}$ is of
order the cumulative Sobolev optical depth along ${\bf {\hat n}}$ summed over all lines.
When raised to the $-\alpha$ power and averaged over the directions
of incident flux,
${\tilde t}$ incorporates all opacity effects and gives the
self-shadowing reduction of the line force.

To connect with standard nomenclature, 
${\tilde t}$ is the CAK parameter $t$ multiplied by
$Q c/v_{th}$, and unlike
$t$ itself, it is independent of both $v_{th}$
and the Sobolev length.
Indeed, neither of these latter parameters
need even be specified here (as long
as they are small), which circumvents certain ambiguities in the 
more standard notation.

\subsection{The Vector Correction for Self-Shadowing}

One particularly convenient way to incorporate the effects of line
optical depth is to
consider the ratio of the radiative force ${\bf {\vec g}}$ 
to $g_{thin}$, the
magnitude of the force if all lines were optically thin (obtained
by setting ${\tilde t} = 1$ in eq. [\ref{gradgeneral}] and considering
only the magnitude).
This ratio gives a vector result with a magnitude less than unity (often
much less), which can be termed 
the vector correction for line shadowing and written as
\beq
{\bf {\vec P}} \ = \ {{\bf {\vec g}} \over g_{thin}} \ .
\eeq
In the absense of strict spherical symmetry, ${\bf {\vec P}}$ is in general
a {\it nonradial} vector.
This holds even when the flux {\it is} radial;
all that is required is an asymmetry in effective opacity caused by
nonspherical flows, such as will be considered below.

To simplify the angular integrals for the purposes of this paper,
we will assume a spherical star with no limb- or gravity-darkening,
despite the likely possibility that both occur in reality.
Also, the disk is assumed to be optically thin in the continuum, and
photons are assumed to interact with at most one line, so the disk
thickness and density are limited.
With these
assumptions, using equation (\ref{tildenq}) with an exponential cutoff at 
$q = Q$ (simply for purposes of defining $g_{thin}$) yields
\beq
\label{vecp}
{\bf {\vec P}} \ = \ {1 \over \pi}
{r^2 \over R^2} \int_{\mu_*}^1 d\mu \int_0^{2\pi}
d\phi \ {\bf {\hat n}}(\mu,\phi) {\tilde t}(\mu,\phi)^{-\alpha} \ .
\eeq
Here $\mu$ and $\phi$ give respectively the direction cosine 
from the radius vector and the
incident azimuthal angle of the photon,
and $\mu_* = \sqrt{1-R^2/r^2}$ gives the direction cosine to the limb.

In applications, it is often convenient to express the radiative force
relative to gravity.
This is normally done in terms of the Eddington parameter $\Gamma$, 
the ratio of the radiative force
on free electrons to gravity (and is not to be confused with the Gamma function
$\Gamma(\alpha)$).
It then follows that
\beq 
\label{vecg}
{\bf {\vec g}} \ = \ 
g_{grav} \Gamma \ {g_{thin} \over g_{es}} \ {\bf {\vec P}} \ ,
\eeq
where the ratio of the radiative force on lines, without any shadowing
correction, to the radiative force on free electrons is by our convention
\beq
\label{goverg}
{g_{thin} \over g_{es}} \ = \ Q \Gamma(\alpha) \ .
\eeq
Thus
\beq
\label{myforce}
{\bf {\vec g}} \ = \ g_{grav} \Gamma Q \ \Gamma(\alpha) \ {\bf {\vec P}} 
\eeq
provides a general and convenient expression for the radiative force
for a statistical distribution of marginally optically thick lines in the
Sobolev approximation.

A key assumption here is that equation (\ref{tildenq}) applies
universally over the 
circumstellar medium, and 
so may be viewed as a global constraint.
Radially dependent effects, such as 
might be linked to gradients in ionization
or temperature, are not explicitly included.
By contrast, Chen \& Marlborough (1994) neglected
the optical depth effects controlled by the 
${\bf{\vec P}}$ vector, but instead allowed the line parameters
to be radially varying in a parametric way,
albeit without clear physical justification.
Although a complete treatment must accomodate
both shadowing by line optical depth and the
radial gradients for an
{\it actual} line list, for conceptual simplicity we choose to include only the
former, because of the importance of the
dynamical dependences it implies.
Also, it is implicitly assumed that there are a large
number of optically thick lines and a statistical treatment is justified, which
also requires that $P_r \<< 1$.
This is more likely to hold in a dense disk than in the weakest B-star winds
(Babel 1996).

\subsection{The Line-of-Sight Velocity Gradient 
in the Equatorial Plane}

The above results are general and 
entirely equivalent to the standard CAK formalism with a fixed line-list
parametrization.
Since our present interest is orbiting disks, we will now consider
the explicit restriction of these equations to that context.
For simplicity we consider only forces in the equatorial plane.
Then the dependence on the line-of-sight gradient of
the line-of-sight velocity becomes (e.g., Batchelor 1967; Cranmer \& Owocki 1995)
\beq
\left | \left ({\bf {\hat n}} \cdot \gradient \right ) 
\left ( {\bf {\hat n}} \cdot
{\bf {\vec v}} \right ) 
\right |\ = \ \left | H(\mu,\phi) \right |
\eeq
where
\beq
\label{hmu}
H(\mu,\phi) \ = \  \mu^2 {\partial v_r \over \partial r} \ + \
(1-\mu^2) {v_r \over r} \ + \ \mu \sqrt{1-\mu^2} \sin \phi \left (  
{\partial v_\phi \over \partial r} \ - \ {v_\phi \over r} \right )  \ ,
\eeq
and the correction for line shadowing becomes
\beq
{\tilde t}^{-\alpha} \ = \ 
\left (Q \rho \kappa_{es} c \right )^{-\alpha} 
\left | H(\mu,\phi) \right |^\alpha \ .
\eeq

\subsection{Radial Forces on Circular Orbits}

Interactions between gas parcels favor orbital circularization.
If circular orbits are assumed, the above expressions simplify considerably,
since then 
\beq
\label{kepler}
v_\phi = \sqrt{{GM \over r}}
\eeq 
and so $\hbox{d}v_\phi/\hbox{d}r =  -0.5 \  v_\phi/r$ may
be substituted into equation (\ref{hmu}).
The negative sign implies that the Keplerian velocity shear {\it augments}
the gradient due to orbital curvature, thereby strengthening the magnitude
of the effects we consider.

Setting $v_r = 0$ for circular orbits then gives
\beq
\label{tcircle}
{\tilde t}({\bf {\hat n}})^{-\alpha} \ = \
\left ({3 \over 2} \right )^\alpha \left ( Q \rho \kappa_{es} c
\right )^{-\alpha} \left ( GM \right )^{\alpha/2} r^{-3\alpha/2}
\mu^\alpha
(1-\mu^2)^{\alpha/2} |\sin \phi|^\alpha \ .
\eeq
From equation (\ref{vecp}) this in turn gives for the radial component
of the self-shadowing correction
\beq
\label{prcircle}
P_r \ = \ {2 \over \pi} \left ({3 \over 2} \right )^\alpha R^\alpha
\left (Q \kappa_{es} c \right )^{-\alpha} \left ( GM \right )^{\alpha/2}
\rho^{-\alpha} r^{-5\alpha/2} A_\alpha f_0(r) \ ,
\eeq
where we have defined for convenience the 
weakly varying order-unity expressions
\beq
\label{aalpha}
A_\alpha \ = \ \int_0^{2\pi} d\phi \ |\sin{\phi}|^\alpha
\eeq
and
\beq
\label{fl}
f_l(r) \ = \ 
{1 \over 4 }  \int_0^1 dx \
x^{\alpha/2} \left ( 1 \ - \ {R^2 \over r^2} x \right )^{\alpha/2 \ - \ l } \ . 
\eeq

Figure 2 plots the quantities $A_\alpha f_0$ and $A_\alpha (1.5  - f_1/f_0)$ 
(the latter
will be needed below in \S 4), expressed
as functions of $r/R$ for selected $\alpha$, to demonstrate their
weak dependence on $r$.
The figure shows that $A_\alpha f_0(r)$ is generally close enough to unity that
conceptual results can safely omit this factor.
The primary dependences are instead contained in the other factors 
in equation (\ref{prcircle}).
Specifically, if the disk density falls steeply as $\rho \propto r^{-n}$,
then the radial radiative force relative to
gravity scales asymptotically like $P_r \propto r^{(n-2.5)\alpha}$.
In this context it is relevant that $n \gtwig 2.5$ is often inferred from
observations (e.g., Waters 1986),
so that the radial radiative force may
fall off similarly to gravity, as occurs when $n = 2.5$ since then $P_r$ is
roughly constant.
It may also fall off less steeply if $n > 2.5$, since then $P_r$ increases with radius.
Note that
the latter situation cannot persist to arbitrarily large radii or the disk
would become unbound far enough from the star.

\begin{figure}
\unitlength1in
\begin{picture}(6.5,5.)
   \leavevmode
   \epsfxsize=6.5in
   \epsffile{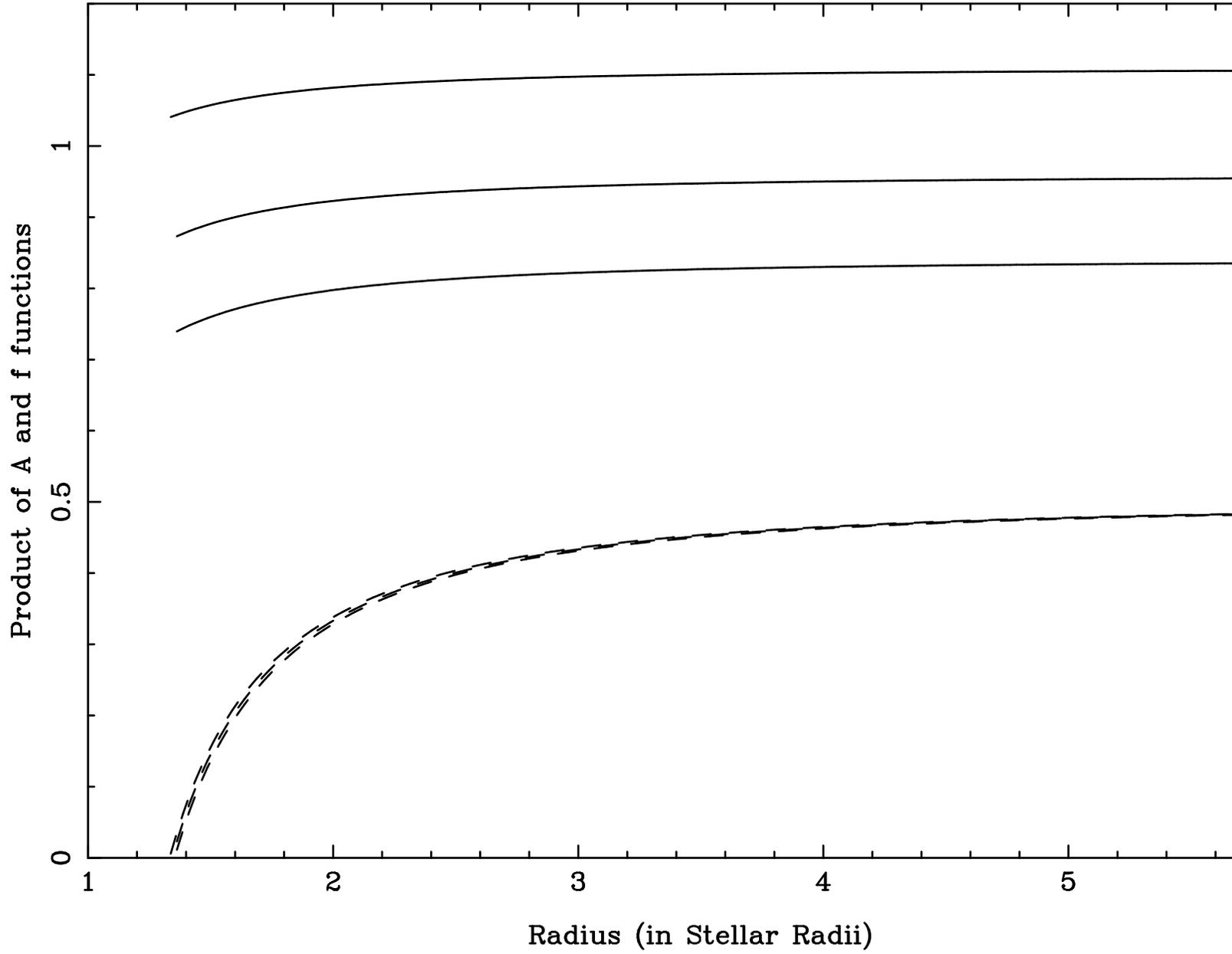}
\end{picture}
\hfill
\caption[ ]{\footnotesize
The radial dependence of the quantities $A_\alpha f_0(r)$ (solid curves)
and $3/2 \ - \ f_1(r)/f_0(r)$ (dashed curves), plotted
for $\alpha = 1/3, 1/2,$ and 2/3.
In all cases, increasing
curve height corresponds to increasing $\alpha$.
The dashed curve is only plotted in its positive regime, where
the radiative torque has the same sign as $v_r$.
\label{fig2}
}
\end{figure}

Thus we stress the surprising and important result that radiation that
is neither radial nor azimuthal
($\mu$ equals neither 0 nor 1) induces CAK-type
line forces due to the orbital gradients (recall Figure 1).
Even in the absence of radial flow,
incident radiation that is 
prograde/retrograde symmetric (i.e., symmetric in $\phi$)
results in a force pointing radially outward.
Note that
since we assumed Keplerian orbits in the derivation of equation (\ref{tcircle}),
which is not strictly consistent with the presence of a radiative force, we
can only use our expressions to test when the 
magnitude of the radiative force will represent
a {\it small} perturbation to gravity.
We quantify this comparison next.

\section{Estimating the Magnitude of Radiative Forces on Circular Orbits}

In the previous section, it was shown that acceleration
by line forces is strongest
when the density is low and self-shadowing is minimized.
The first point of interest is to determine
how low the density needs to be to enable the
radiative force to compete with gravity
and alter the Keplerian structure.
Thus our approach of expressing the radiative acceleration as a ratio
to gravity will continue to be convenient.

\subsection{Radial Radiative Acceleration Compared to Gravity}
 
To quantify this ratio, we apply equation (\ref{myforce}) 
subject to
equation (\ref{prcircle}).
The total radial line force in units of gravity is then
\beq
\label{gtograv}
{g_r \over g_{grav}} \ = \ 
\Gamma(\alpha) {2 \over \pi} \left ({3 \over 2} \right )^\alpha
A_\alpha f_0(r)
\Gamma Q^{1-\alpha}
(G M)^{\alpha/2}
R^\alpha \left ( \rho \kappa_{es} c \right )^{-\alpha}
r^{-5\alpha/2} \ .
\eeq
This result, along with
equation (\ref{kepler}),  shows that the overall scale of the radiative force in units
of gravity is set by
$\Gamma Q$ times the factor $(Q \rho \kappa_{es} r c/ v_\phi)^{-\alpha}$,
which is quite similar to the wind result with the characteristic velocity
replaced by $v_\phi$.

Gayley (1995) argues that $Q$ may canonically be of order $10^3$.
A careful analysis by Puls {\it et al.} (2000) 
indicates that, for effective temperatures characteritic of B-stars,
the meaning of the $Q$ parameter requires generalization.
Taking their results for $T_{eff} = 20,000 K$
allows 
us here to use an  
an effective value of $Q$,
given by $Q \ =  \ {\bar Q}^{1/(1-\alpha)}Q_o^{-\alpha/(1-\alpha)} \ = \ 315 $
with $\alpha = 0.58$, 
although it is not clear how the higher densities characteristic of disk
might alter this result.
In any event, for line-driven wind theory to be valid, the product
$\Gamma Q$ must substantially exceed unity.
If we assume a nominally bright type of
Be star, similar to the B0V model used in
Okazaki (2000), with $\Gamma = 5 \times 10^{-2}$ and $R = 5 \times 10^{11}$ cm, 
and a disk with rough
parameters $\rho \sim 10^{-12}$ g cm$^{-3}$, 
$\kappa_{es} \sim 0.3$, 
and $v_\phi \sim 4 \times 10^7$ cm s$^{-1}$, then equation (\ref{gtograv}) gives 
that a crude estimate of the force ratio is of order $10^{1-6\alpha}$ at 2 stellar radii.
This is quite sensitive to $\alpha$, but may be of order several percent
if $\alpha$ drops below 1/2, and possibly tens of percent if $\alpha$ drops
below 1/3.
The ratio can even approach order unity if a low $\alpha$ is
combined with high $\Gamma$, high $Q$, and
low $\rho$, so the potential exists for the basic structure of some Be disks
to be substantially altered by optically thick line forces.

The above quantification is admittedly crude, since the
appropriate assumptions in a high-density disk environment
are as yet uncertain, and the line-list parameters are left here as
free parameters that await future clarification.
Still, the results suggest that
the radiative force may at times
represent an important perturbation to
the gross disk structure,
and consistency checks are advisable in general.
This is particularly crucial in regions of lower
density, such as would be found 
many scale heights above the
disk midplane, or many stellar radii out into disks with $n > 2.5$.
The former effect could have important ramifications for the
lower boundary that supplies the mass and angular momentum for disk winds,
while the latter could impose an outer limit to the disk radius.
Furthermore, even when the radiative 
force is found to be marginally negligible for the
overall disk dynamics, it will have magnified importance for
higher-order disk phenomena, such as
the one-arm modes discussed by Okazaki (1997, 2000)
and others (Papaloizou, Savonije, \& Henrichs 1992), 
a point to which we will return in \S 4.



\subsection{Radial Radiative Acceleration in Disks Relative to Winds}

An alternative approach for characterizing the overall magnitude of the
radial radiative acceleration
in disks is to compare it to the acceleration of the
polar stellar wind.
This approach is informative because if we
assume the line-list parameters in the disk are similar to those in
the wind, then both the unknown line-strength parameter $Q$ and the
stellar luminosity affect the disk acceleration 
in
the same way as the wind acceleration, and so cancel in the ratio.
Also, other unknowns can be subsumed into the observed density ratio between
the disk and the wind.

Some of these assumptions may be questionable, since the 
the higher disk
density should cause a lower degree of ionization,
and this is likely to {\it increase} the efficiency of the radiative coupling,
due to the increased number of lines.
In that case, we obtain only a lower limit for
the radiative force on the disk, so the conclusions of this pilot study
may actually represent {\it underestimates} of the effect.
Notwithstanding these uncertainties, the disk/wind radiative
acceleration ratio provides a useful reference, because 
without including the complications endemic to weak winds
(Babel 1996), the polar wind
acceleration is
also constrained relative to gravity by CAK theory.

To make this comparison, we must express ${\tilde t}$ in the wind,
which requires an assumption about the velocity structure.
For simplicity we assume the wind velocity is proportional to 
$1 \ - \ R/r$, which is often termed a ``$\beta$ = 1'' law.
This implies that for a nonrotating wind, equation (\ref{tcircle}) is replaced by
\beq
\label{twind}
{\tilde t}({\bf {\hat n}})^{-\alpha} \ = \ \left (
c Q \rho \kappa_{es} \right )^{-\alpha} \left [1 \ - {R \over r} \ + \
\mu^2 \left ( 2 {R \over r} - 1 \right ) \right ]^\alpha \left (
{v_{\infty} \over r} \right )^\alpha \ .
\eeq
Then we can write for the radial radiative acceleration ratio
\beq
{g_{D} \over g_{W}} \ = \ 
{P_D(r) \over P_W(r)} \ ,
\eeq
where the $W$ and $D$ denote wind and disk respectively.
Evaluating $P_D$ and $P_W$ using equation (\ref{vecp}) 
with equations (\ref{tcircle}) and (\ref{twind})
respectively then yields
\beq
\label{grdovergrw}
{g_{D} \over g_{W}} \ = \ 
\xfun(r) \left [ {\rho_W(r) \over \rho_D(r)}
\right ]^\alpha 
\eeq
where we have defined the function
\beq
\xfun(r) \ = \ 
{2 \over \pi}
\left ( {1 - \alpha \over 2.1 \alpha} \right )^\alpha (1+\alpha)
A_\alpha f_0(r) 
{R^{\alpha/2+1} (2R-r) r^{3\alpha/2} \over [r^{2+2\alpha} -
(r^2+rR-2R^2)^{1+\alpha}]} 
\eeq
which encodes all the geometrical differences between Keplerian disks and 
radially streaming winds with $\beta = 1$.
Here it has further been assumed that the stellar wind terminal speed
can be characterized by the empirical relation 
$v_\infty/v_{esc} \cong 2.2 \alpha/
(1-\alpha)$ (approximated roughly from the analytic solution of
Kudritzki {\it et al.} 1989), and neither limb- nor
gravity-darkening are included.

The function $\xfun(r)$ is plotted in Figure 3 for $\alpha = $ 1/3, 1/2, and 2/3.
Note that the decrease in $\xfun$ with $\alpha$
arises partly from the more rapid wind acceleration
for high $\alpha$, and partly from the details of
the radiative force in the disk.
More fundamentally, and contrary to
the expectations (e.g., Okazaki 2000), since $\xfun(r)$ is {\it not} small the
magnitude of
$g_D/g_{W}$
is controlled primarily by the $(\rho_W/\rho_D)^\alpha$ factor.
This expresses the fact that the
self-shadowing correction reduces the acceleration substantially only if the
disk has a much higher density than the wind, and, especially for low $\alpha$,
the geometrical differences between a Keplerian disk and a wind are not as
large as might be expected.
Thus the primary difference between CAK-type forces in a disk as opposed to
a wind, in the single-scattering limit for the UV continuum, arises from
the density difference and not the geometrical differences, and
this is especially true for low $\alpha$.
Furthermore, when $\alpha$ is low,
weaker lines dominate and self-shadowing is less severe, so even
the density difference has a less pronounced impact.

\begin{figure}
\unitlength1in
\begin{picture}(6.5,5.)
   \leavevmode
   \epsfxsize=6.5in
   \epsffile{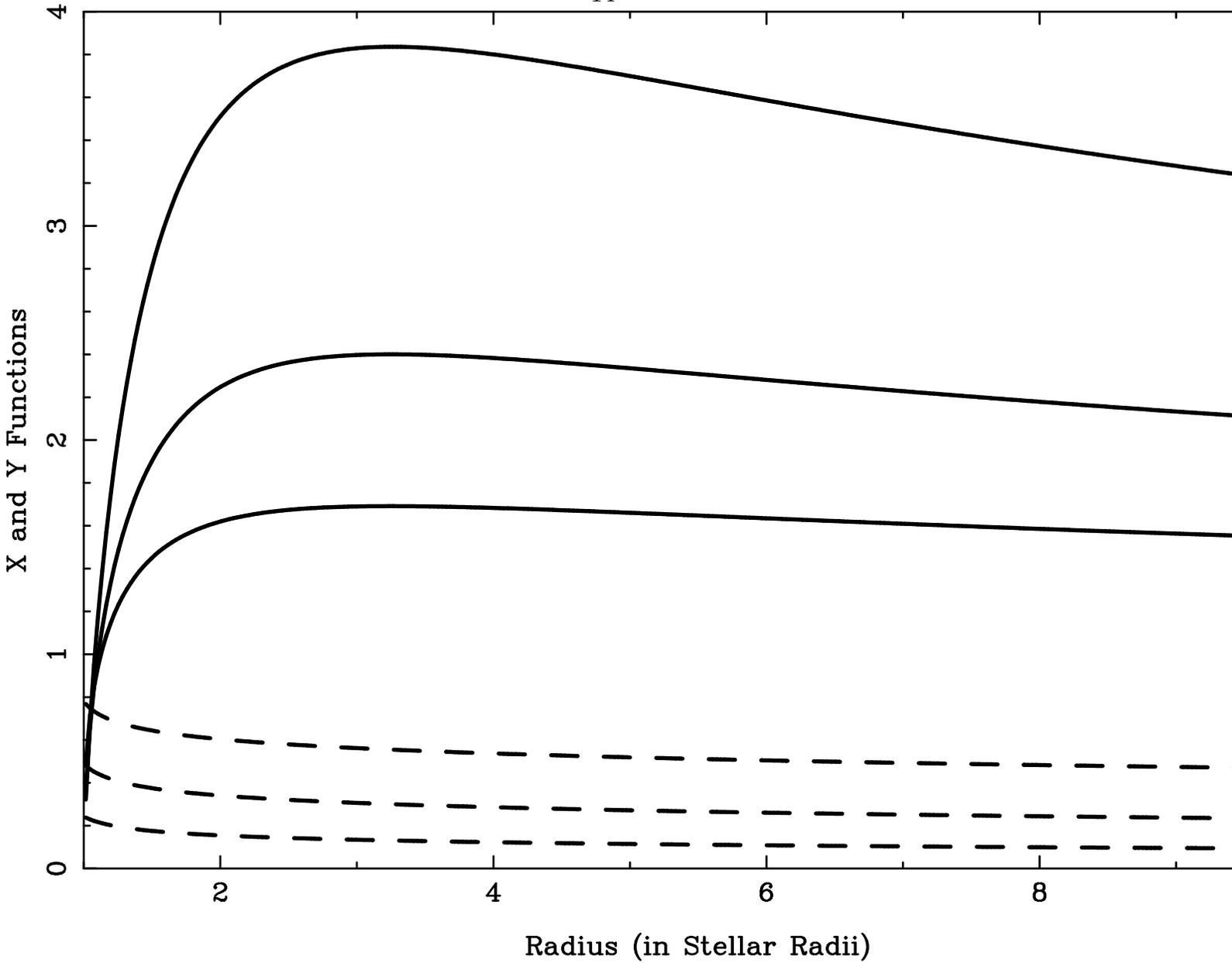}
\end{picture}
\hfill
\caption[ ]{\footnotesize
The radial dependence of the geometric functions $X(r)$ (dashed curves; the
ratio of disk acceleration to wind acceleration at the same density)
and $Y(r)$
(solid curves; the ratio of disk acceleration to gravity assuming windlike
properties)
are plotted, again for $\alpha = $1/3, 1/2, and 2/3.
The lower values of $X(r)$, and the higher values of
$Y(r)$, are associated with higher $\alpha$.
(The reversal is due to the fact that $X(r)$ is influenced by the
faster wind acceleration when $\alpha$ is high.)
\label{fig3}
}
\end{figure}

In summary, when $\alpha$ and the disk density are both high, it is clear that
the radiative acceleration in the disk
drops significantly.
On the other hand, for low $\alpha$,
regions of the disk which have densities comparable
to the polar wind, such as may occur near its edges or at large
radii,
will experience a radiative acceleration that is of the same order as
in the wind.
We may even
conclude from Figure 3 that for sufficiently small values of $\alpha \ltwig 1/3$,
radiative accelerations
that are of the same magnitude as those in the wind
may be achieved even at depths in the disk where
the density exceeds the wind density by two orders of magnitude.

The appearance of windlike accelerations at the edge of the disk is the cause
of a disk wind, so these results suggest
there may be a more gradual transition from a Keplerian
disk into a disk wind than is generally assumed
(see e.g., Proga, Stone, \& Drew 1999;
Feldmeier, Shlosman, \& Vitello 1999).
If so, the trading of centrifugal support for radiative support could
reduce the specific angular momentum, and
induce orbital shears and angular momentum
transport in the boundary layer
below the disk wind, in ways that are
yet to be quantified.
Future work on the effects of radiative forces on steady-state
disk dynamics is planned, to elucidate the anticipated
ramifications for radiatively driven 
mass-loss rates and the critical-point topology.

Furthermore, as mentioned above, we may actually be underestimating the
forces in the 
denser, cooler, and less ionized disk by assuming that the line list
is essentially the same in the disk as in the wind.
Therefore,
firm constraints on the expected values of the line-opacity parameters
in disks are required before disk-structure models can safely neglect
the radial radiative force.



\subsection{Relation to Gravity Using the Disk/Wind Density Ratio}

Ultimately, the key is not how the disk acceleration compares to
that in the wind, but rather how it compares to gravity.
As mentioned above, expressing
the radiative acceleration in the disk in terms of the
acceleration in the polar stellar wind allows us
another approach for comparing it to gravity, since 
gravity and the wind acceleration
have in theory a fixed relationship.
If we assume the wind is losing mass at the maximum rate consistent with
the CAK formalism, this relationship is (Friend \& Abbott 1986;
Pauldrach, Puls, \& Kudritzki 1986)
\beq
{g_W \over g_{grav} } \ = \ {(1+\alpha)F \over \alpha^\alpha (1-\alpha)^{1-\alpha}}
\left ( {g_{net} \over g_{grav}} \right )^\alpha \ ,
\eeq
where $g_{net}$ is the {\it actual} net acceleration of the wind and 
$F$ is the ``finite disk'' correction factor, evaluated below.
Note that at the critical point, $F \cong 1/(1+\alpha)$ and 
$g_W/g_{grav} \ = \ 1/(1-\alpha)$.
If we again assume $v_\infty/v_{esc} \ \cong 2.2 \alpha/(1-\alpha)$
with a $\beta = 1$ wind, 
then
\beq
{g_{net} \over g_{grav} } \ = \ 9.68
{\alpha^2 \over (1-\alpha)^2} \left ( 1 - {R \over r} \right )
\eeq
and
\beq
\label{gwograv}
{g_W \over g_{grav} } \ = \ 9.68^\alpha {\alpha^\alpha \over
(1-\alpha)^{1+\alpha}} 
(1+\alpha) F \left ( 1 - {R \over r} \right )^\alpha \ ,
\eeq
where the finite-disk correction factor is
\beq
F \ = \ {r^2 \over (1+\alpha) R (2R - r)} \left \{
1 \ - \ \left [1 - \left ( 2 - {r \over R} \right )
{R^2 \over r^2} \right ]^{1+\alpha} \right \} \ .
\eeq

Equation (\ref{gwograv}), coupled with equation (\ref{grdovergrw}), allows us to
find a new expression for $g_D/g_{grav}$, where many of the variables
in the previous equation (\ref{gtograv}) are replaced by the observable density
ratio between disk and wind.
The result is
\beq
\label{gtogravnew}
{g_D \over g_{grav}} \ = \ 
\yfun(r)
\left [ {\rho_W(r) \over \rho_D(r)} \right ]^\alpha 
\eeq
where
\beq
\yfun(r) \ = \ {2 \over \pi}
4.6^\alpha {(1+\alpha)^2 \over (1-\alpha)} A_\alpha f_0(r) F
{r^{\alpha/2} R^{\alpha/2+1} (r-R)^\alpha (2R-r) \over
\left [ r^{2+2\alpha} - (r^2+rR-2R^2)^{1+\alpha} \right ]} \ .
\eeq
In comparison with equation (\ref{gtograv}), this 
alternate form removes the unknown $Q$ parameter and demonstrates
explicitly the importance of the density ratio,
since $\yfun(r) \gtwig 1$
and varies only gradually with $r$ and $\alpha$, as shown in Figure 3.

The detailed behavior of $\yfun(r)$ is sensitive to the
schematic choice $\beta = 1$ and is not intended to be quantitatively
reliable; only its overall magnitude is significant. 
Note further that it is simply connected to $\xfun(r)$
by $\yfun/\xfun = g_W/g_{grav}$ (and see eq. [\ref{gwograv}]),
so the fact that Figure 3 shows $\yfun(r)$ to be substantially larger
than $\xfun(r)$ is simply due to the fact that a CAK-type wind accelerates
substantially faster than gravity (Gayley 2000).
The point being made here
is simply that if a Keplerian disk had the same
density and line properties as a stellar wind at some $r$, then the
azimuthally orbiting gas would
actually experience a {\it larger} outward radiative acceleration than 
the radially streaming gas.
This would of course invalidate the Keplerian disk
assumption, and suggests that disks around UV-bright stars
must have higher than windlike densities to avoid being rapidly blown off.

Since the $\yfun$ function is seen to vary only over about a factor of
2 over much of the disk for a wide range
in $\alpha$, it is clear that the importance of the $\alpha$ parameter
appears primarily as the exponent of the disk density, relative to
canonical CAK-type winds.
The fact that the disk is expected to be orders of magnitude denser than
the wind indicates that this dependence on $\alpha$ can be quite steep.
It is thus significant that
Puls {\it et al.} (2000) argue in favor of $\alpha \sim 0.3$ in
B-star winds, a fairly low value which is also supported by the low terminal
speeds around these stars.
If such low $\alpha$ is characteristic of the disk
as well, we conclude that even Keplerian disks that are denser
than winds by 2--3 orders of magnitude
may experience line
forces that modify gravity at
the 10--20\% level, and the potential impact on disk models should not be overlooked.
This holds even for purely circular orbits; the subtle ramifications for
radiative forces when
orbital eccentricity appears is treated next.

\section{Precession of Slightly Elliptical Orbits Due to Radiative Forces}

Even when the radiative force is small compared to gravity, it can have
an important cumulative effect over many orbits.
Here we show that even weak radiative forces should be expected to cause
precession of linear perturbations to gas parcels in circular orbit.
However, substantial uncertainly remains because we find
one important effect which gives rise to prograde precession while another of
comparable importance leads to retrograde precession, leaving even the predicted
{\it sign} in question.

\subsection{Radiative Perturbations to First Order in Eccentricity}

If slightly elliptical orbits are allowed, $v_r$ is no longer zero,
and a more complicated expression for equation (\ref{hmu}) becomes  
necessary.
Nevertheless, treating $v_r$ as small compared to $v_\phi$ again results
in a form that is
convenient for analytic manipulation.
To first order in $v_r/v_\phi$ the radial force is unaffected,
but significantly, an 
azimuthal {\it torque} appears.

The presence of this torque is a subtle consequence of the angular character
of Sobolev opacity, which is sensitive to the line-of-sight velocity gradient.
This gradient is azimuthally asymmetric in the presence of rotation and
expansion (Grinin 1978), and the resulting generation of
radiative torque has been studied in the limit
of {\it large} $v_r/v_\phi$ in stellar winds by Gayley \& Owocki (2000).
Here we consider the opposite limit, applicable to orbits with low ellipticity.

If we work to lowest order in $v_r/v_\phi$, then
${\tilde t}({\bf {\hat n}})^{-\alpha}$ has a different action on
even and odd components of the 
azimuthal distribution of incident radiation. 
The appropriate value for ${\tilde t}^{-\alpha}$ is the same as equation (\ref{tcircle}) 
when applied to the {\it even} moment responsible for the radial force, but
for the {\it odd} moment responsible for the azimuthal torque, it becomes
(using eq. [\ref{hmu}]) to lowest order
\beq
\label{todd}
{\tilde t}_{odd}({\bf {\hat n}})^{-\alpha} \ \cong \ 
- \left ({3 \over 2} \right )^{\alpha-1}
\alpha (1 + \sigma \mu^2)\mu^{\alpha-1}
(1-\mu^2)^{(\alpha-1)/2}
{|\sin \phi|^\alpha \over \sin \phi}
\left | {v_\phi \over r} \right |^\alpha {v_r \over v_\phi}
(c Q \rho \kappa_{es})^{-\alpha} \ ,
\eeq
where $\sigma = \partial \ln v_r / \partial \ln r \ - \ 1$.

\subsection{Torque on Aligned Fixed-Eccentricity Orbits}

The value of $\sigma$ requires an assumption
about the orbital geometry.
Note that if $\partial v_r/\partial r \ < \ 0$, then $\sigma \ < \ -1$, and
the $1+\sigma \mu^2$ term changes sign at high $\mu$.
This is an important consideration for the torque, since the
prograde/retrograde symmetry is broken by the competition between
the orbital expansion/compression along oblique 
prograde/retrograde angles, and the expansion
or compression due to $v_r$.
The radiative torque due to angles with 
$1+\sigma \mu^2 \ < \ 0$ has the same sign as $v_r$ and adds angular
momentum when $v_r \ > \ 0$, whereas $1+\sigma \mu^2 \ > \ 0$ removes it.
Thus the value of $\sigma$ controls the sense of the torque, and as we
show below, the sense of the precession.

Since a one-arm mode is an azimuthally 
coherent structure, the major axes of the
embedded elliptical orbits must align.
Furthermore, the density perturbations will peak where the
orbital eccentricity peaks.
When the eccentricity is maximal, it is also locally constant,
describing a local family of homologous orbits with a range of
periastron distances.
For given eccentricity $\varepsilon$, the radial velocity
in nearly circular orbits becomes
\beq
\label{vrphi}
v_r \ \cong \ \varepsilon \sin \phi \sqrt{ G M \over r_p} \ ,
\eeq
where $\phi$ is the orbital angle measured from periastron and
$r_p$ is the periastron distance. 
Since $r$ is close to $r_p$, equation (\ref{vrphi}) gives
\beq
\sigma \ \cong \ -{3 \over 2} \ + \ {d \ln \varepsilon \over d \ln r_p} \ ,
\eeq
so that $\sigma \cong -3/2$ whenever $\varepsilon$ is slowly varying.
Note that when $v_r \ > \ 0$, this also implies $\partial v_r/\partial r  <  0$ (i.e.,
the radial velocities are convergent), despite
the fact that neighboring orbits themselves are divergent.
This counterintuitive behavior occurs because the angular velocity
of neighboring orbits is not the same, so outer orbits lag inner ones
and gas in the outer orbit does not increase in $r$ as
rapidly as gas in the inner orbit, so the radial components of the individual gas
parcels converge even as their orbital traces diverge.

Using $\sigma = -3/2$ and equation (\ref{todd})
in equations (\ref{vecp}) and (\ref{myforce}), we thus find that the (small) azimuthal 
acceleration owing to the radiative torque is
\beq
\label{torque}
g_\phi \ = \ g_{grav} \Gamma Q \Gamma(\alpha) P_\phi
\eeq
where
\beq
\label{pphi}
P_\phi \ = \ 
{2^{2-\alpha} 3^{\alpha-1} \over \pi} \alpha A_\alpha \left [
{3 \over 2} f_0(r) - f_1(r) \right ] (Q \kappa_{es} c \rho)^{-\alpha}
\left ( {R \over r} \right )^\alpha \left | {v_\phi \over r} \right |^\alpha
{v_r \over v_\phi} \ .
\eeq
Note that the torque is proportional to
$v_r$, 
a convenient property we will exploit below.
Like the coriolis force in the nearly corotating frame,
this has the important ramification that the
sign of the torque reverses when $v_r$ reverses, so
the torque is applied antisymmetrically and thus imparts no 
net angular momentum over a
complete cycle between turning points.
In the presence of this torque the
orbits are no longer strictly elliptical,
and indeed do not close on themselves, but
the energy and angular momentum become 
periodic functions of $r$, so $r(t)$ remains periodic.
The small deviation from Keplerian motion can then only yield orbital {\it precession}.

The assertion that
the motion is symmetric about the turning points follows from the fact
that the force
depends only on radius and velocity and is
invariant under time reversal.
Thus bounded motion must progress from turning point to turning point in
the same way regardless of whether
time runs forward or backward, so there is no hysteresis and $r(t)$ is periodic.
The time reversibility is obvious for the {\it magnitude}
of optically thick radiative forces, which
in the Sobolev approximation depend only on the magnitude of the
line-of-sight velocity gradient.
This invariance also applies to the
{\it sign} of the radiative torque, because equation (\ref{torque}) shows that
reversing the radial
velocity $v_r$
changes the sign of the torque {\it relative} to the azimuthal velocity
$v_\phi$,
which is itself reversed.

The lack of hysteresis between
turning points does not imply that the ``orbits'' are closed, because the
azimuthal angle covered in each full cycle is
not necessarily equal to $2 \pi$.
Instead, in the presence of torque,
the apsides will precess.
Before calculating this precession, we first quantify the radiative 
torque magnitude by comparing it to 
the lateral force that is responsible
for maintaining closed orbits in the rotating frame, namely the coriolis force.

\subsection{Magnitude of the Radiative Torque Relative to the Coriolis Force}

The coriolis force in the corotating frame of a 
nearly circular orbit gives a fictitious azimuthal acceleration
\beq
g_{cor} \ = \ -2 \Omega_p v_r
\eeq
for $\Omega_p$ the orbital angular velocity at periastron for nearly
circular Keplerian orbits.
Note that $g_{cor}$, like the radiative $g_\phi$, is proportional to $v_r$,
so a simple ratio exists,
which from equation (\ref{todd}) applied in
equation (\ref{vecp}) is given by
\beq
\label{overn}
{g_{\phi} \over g_{cor}} \ = \ - \Gamma(\alpha) \alpha 
\left ( {3 \over 2} \right )^{1-\alpha}
A_\alpha \Gamma Q \left ( {| v_{\phi} | \over c Q \rho \kappa_{es} r}
\right )^\alpha 
\sqrt{{r \over R}} \int_{\mu_*}^1 d\mu \ \mu^{\alpha-1}
(1-\mu^2)^{\alpha/2} \ .
\eeq

\subsection{Precession Due to Torque Proportional to Radial Speed}
 
We wish to quantify the precession of a homologous (constant eccentricity) family of
slightly elliptical orbits of gas parcels, all with the same major
axis direction so the family yields a coherent density perturbation.
It is not necessary to solve the equations of motion; the behavior
is specified by solving for
the energy and angular momentum alone.
Note that here these quantities are not constant within a given orbit
because of the work and the torque applied by the radiative force.
 
It is convenient to
write equation (\ref{torque}) in the simple form
\beq
\label{gphisimple}
g_\phi \ = \  {\Omega \over N_\phi} v_r \ ,
\eeq
where $\Omega = v_\phi/r = \sqrt{GM/r^3}$ is the angular velocity of the orbit to
lowest order.
Here the key parameter
\beq
\label{nphitorque}
N_\phi \ = \  {\pi \over \Gamma Q^{1-\alpha} \Gamma(\alpha) 2^{2-\alpha}
3^{\alpha-1} \alpha A_\alpha \left [ {3 \over 2} f_0(r) - f_1(r) \right ]}
(c \kappa_{es} \rho)^{\alpha} \left ( {r \over R} \right )^\alpha
\left ( {r^3 \over GM} \right )^{\alpha/2}
\eeq
remains approximately constant
over nearly circular orbits, and the Appendix shows that it can be
interpreted as the number of 
nearly-closed orbits
required for a full precession cycle, due to the effects of the torque alone.
If its sign is positive, precession is prograde, and if negative, retrograde.
This distinction depends on
$r$ and $\sigma$, but in typical cases of interest
($r \gtwig 2R$, $\sigma \cong -1.5$),
the 
radiative torque yields positive $N_\phi$ and contributes to prograde precession.

To estimate the magnitude of the
precession rate, we again consider $Q = 315$ and $\alpha = 0.58$,
and adopt similar parameters for a B0V star as used by Okazaki (2000) with
$\Gamma = 4.5 \times 10^{-2}$,
$M= 17.8$ M$_\odot$, $r = 5.2 \times 10^{11}$ cm,
and a disk density of $\rho = 10^{-12}$ gm cm$^{-3}$.
With equation (\ref{nphitorque}), these yield 
$N = 637$ at
roughly 3 stellar radii, corresponding to a precession period of 5.3 years.
Observed precession periods are also of order a few years (Mennickent,
Sterken, \& Vogt 1997), and 
the few whose orbital sense is known are all prograde,
so this suggests that radiative torque may be an important contributor.
However, we will show next that a competing effect also exists which may
act to reverse the sign of radiatively induced precession.

\subsection{Precession Caused by Non-Inverse-Square Radial Acceleration}

The presence of slight ellipticity not only induces torques, it implies
that $r$, and therefore the radiative force,
changes over the course of an orbit.
If such changes differ from an inverse-square force law,
they also cause precession, in a manner similar
to a quadrupole moment of gravity (Papaloizou, Savonije, \& Henrichs 1992).
To discover the magnitude of this effect, it is 
convenient to write the radial radiative force in the form
\beq
\label{nrform}
g_r \ = \ g_{grav} {2 \over (2-m_{rad})} \left (
{r \over r_p}\right )^{-m_{rad}} 
{1 \over N_r} \ ,
\eeq
where $r_p$ is the periastron radius and 
\beq
m_{rad} \ = \ {d \ln g_r \over d \ln r}
\eeq
governs the gradient of $g_r$ {\it over a single elliptical orbit} (i.e.,
for a fixed periastron distance $r_p$).
In the Appendix,
it is shown that $N_r$ may be interpreted as the number of orbits per
precession cycle, due solely to the gradient in $g_r$ over an orbit.
Note that negative values for $N_r$ imply retrograde precession
and occur whenever $m_{rad} > 2$.

The determination of $m_{rad}$ depends on the 
density gradient over one orbit, as well as on the assumed
eccentricity gradient, since the latter
thus controls the 
gradient in velocity between neighboring orbits.
The simplest assumption is that neighboring orbits have the 
same eccentricity, which is applicable near eccentricity extrema.
Assuming that such extrema are also relevant to the behavior of
global modes allows our results to be suggestive for global mode
precession as well, although a self-consistent calculation is needed.
Furthermore, we assume the lines of nodes of all orbits are aligned,
so similar elliptical orbits are embedded in a simple way to form
our approximation to the behavior of a global mode.

If it is also assumed that the temperature does not
vary, fixed eccentricity implies that $\rho \propto r^{-1}$ over an orbit,
because conservation of angular momentum implies a compression that
is proportional to $r$, which cancels the $r^{-1}$ dependence from
the stretching
in the vertical direction due to the rising scale height, which leaves
only the radial $r^{-1}$ stretching due to the similarity of the nested
ellipses.
Meanwhile, the remaining geometric effects
involving the flux integral over the line-of-sight velocity gradient indicates
that $g_r \propto \rho^{-\alpha} r^{-5\alpha/2}$, if we
neglect the weak variations in $f_0(r)$ (see eq. [\ref{prcircle}]).
Thus we obtain for aligned orbits with constant eccentricity
\beq
m_{rad} \ = \ 2 \ + \ {3 \alpha \over 2} \ ,
\eeq
and so equation (\ref{nrform}) with $g_{grav} = r \Omega^2$ 
implies
\beq
\label{nrfinal}
N_r \ = \ 
-{4 \over 3 \alpha} 
{r \Omega^2 \over g_r} 
\eeq
gives the number of 
orbits per (retrograde) precession cycle,
due solely to the radiative force gradient.
This expression may be callibrated by comparison with eq. (\ref{gtograv}).

\subsection{Degree of Cancellation of Radiative Precession Effects}

We have seen that at most radii, the torque effect by itself would yield prograde 
precession, while the radiative gradient effect would yield retrograde.
Since these two effects are of similar order, they compete, and a substantial
degree of cancellation occurs.
How this would affect a coherent one-arm mode has yet to be calculated, but
in the situation considered here, that of aligned fixed-eccentricity orbits
with no gas pressure, it is straightforward to show that the gradient effect
is the larger of the two.
This can be seen by considering the ratio $N_\phi/N_r$, which due to
equations (\ref{gphisimple}) and (\ref{nrfinal}) can be written
\beq
{N_\phi \over N_r} \ = \ -{4 \over 3 \alpha} {P_\phi v_\phi \over
P_r v_r} \ .
\eeq
Then equations (\ref{prcircle}) and (\ref{pphi}) further imply
\beq
\label{nratio}
{N_\phi \over N_r} \ = \ -{8 \over 9} \left ( {3 \over 2} \ - \ 
{f_1(r) \over f_0(r)} \right ) \ .
\eeq
An asymptotic analysis reveals that in the limit of large $r/R$, this
becomes
\beq
\label{nratasymp}
{N_\phi \over N_r} \ \cong \ -{4 \over 9} \left [ 1 \ - \
{(4+2\alpha) \over (4+\alpha)} {R^2 \over r^2} \right ] \ .
\eeq
Thus $N_\phi/N_r$ has little $\alpha$ sensitivity at large radii,
stemming from the weak $\alpha$ sensitivity of $f_1(r)/f_0(r)$.
The competition between prograde and retrograde precession has
an intrinsic character that is nearly independent of the line distribution.

Since the smaller the precession number $N$ the stronger the effect,
equation (\ref{nratasymp}) implies 
that at large radii the gradient effect is marginally dominant, yielding
a net retrograde precession.
Of course, for $r \ltwig 1.3 R$ where the dashed curves in Figure 2 go 
negative, the torque effect also yields
retrograde precession.
Thus the results here predict retrograde precession
is always the result of line forces in the limit of small perturbations from
circular orbits, when the orbital eccentricities are co-aligned and constant.

The quantitative results for $N_\phi$ and $-N_r$, converted into timescales
by multiplying by the orbital periods for the characteristic stellar
parameters listed above, are independently graphed in Figure 4.
The ratio between the curves can be seen to agree with equation (\ref{nratio}),
and the magnitudes are seen to roughly correspond to yearlike timescales.
Also plotted is 
the result for the quadrupole precession effect on similar orbits for
the parameters in Papaloizou, Savonije, \& Henrichs (1992), where it can be
seen that radiative effects may compete quite effectively with the 
prograde precession caused by gravitational
perturbation.

\begin{figure}
\unitlength1in
\begin{picture}(6.5,5.)
   \leavevmode
   \epsfxsize=6.5in
   \epsffile{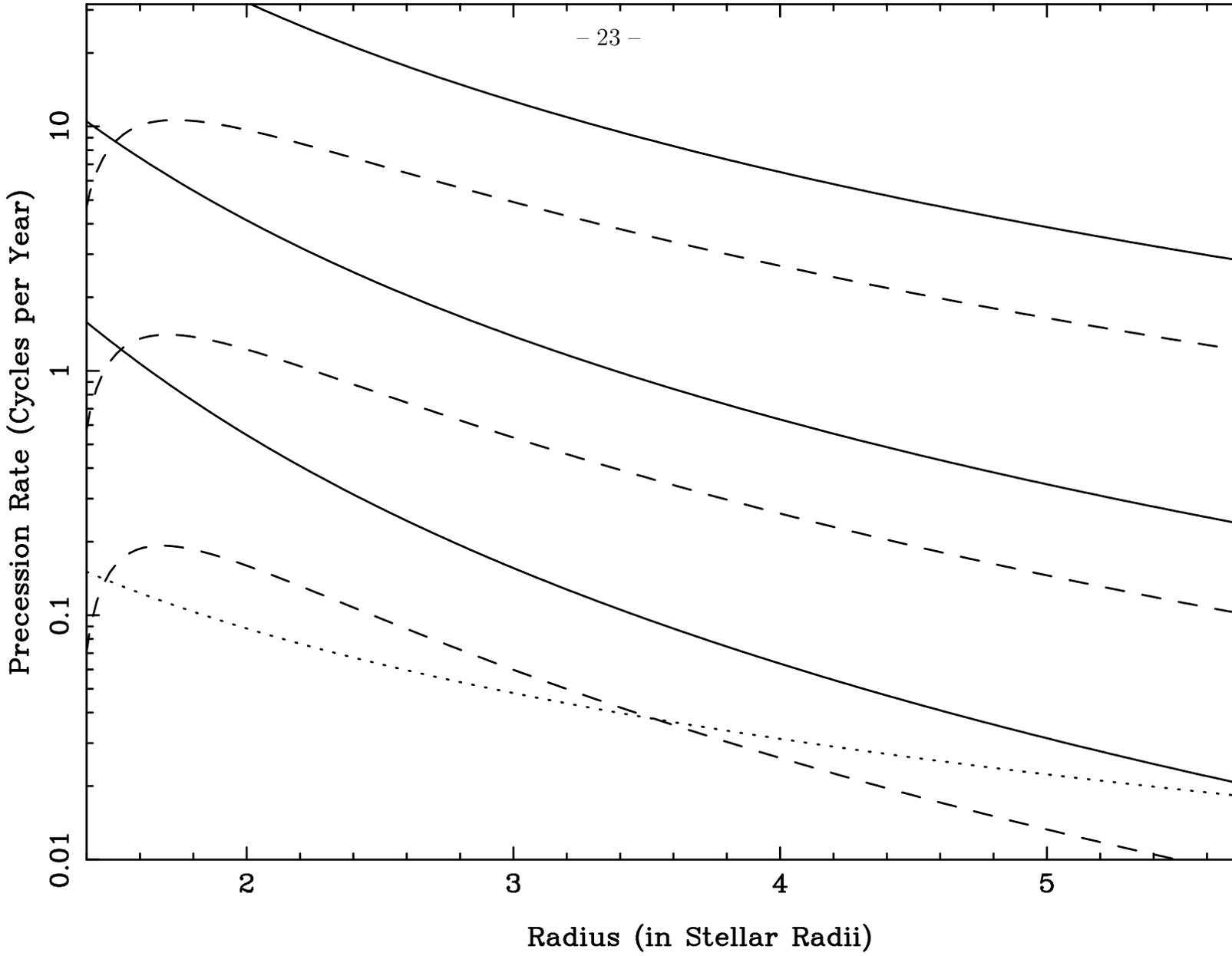}
\end{picture}
\hfill
\caption[ ]{\footnotesize
The rates of retrograde precession due to the gradient effect (solid
curves) and prograde precession due to the torque effect (dashed curves)
are plotted as functions of radius of the individual orbits,
for the stellar parameters in the
text, again using $\alpha = $ 1/3, 1/2, and 2/3.
In both the solid and dashed
sequences plotted, faster precession rates correspond to lower $\alpha$.
Also plotted for comparison is the prograde precession rate due to
the quadrupole correction to gravity (dotted curve)
taken from the model parameters
in Papaloizou, Savonije, \& Henrichs (1992).
As shown in the Appendix, the combined precession rate is found by simply adding the
contributions, where retrograde rates (solid curves) are negative.
There is no global mode present, the rates are for individual slightly elliptical orbits
with no gas-pressure coupling.
\label{fig4}
}
\end{figure}

Although this 
analysis does not guarantee that a coherently growing mode including
pressure forces would also show retrograde precession, it suggests that
the radiative effects on such a mode would include two separate and competing
terms of similar magnitude, and retrograde precession seems a possible outcome.
That observations typically show prograde precession suggests that either the
quadrupole correction to gravity dominates (see the Appendix), or that the
radiative effects in a one-arm mode are not well modeled by small fixed-eccentricity
modifications to circular orbits.
One suggestion that the latter could be true is that the observed amplitudes of
one-arm modes (Hummel \& Hanuschik 1997) 
are often not small, and would fall well outside the
linear regime treated here.
Future nonlinear calculations are needed, perhaps advancing on the analytic work
of Lee \& Goodman (1999).
The magnitude of the results here
indicate that optically thick line forces should be included in any future such effort.

\section{Conclusions}

This paper argues that partially optically thick
line forces are a potentially important contributor to the force balance
in Keplerian disks orbiting Be stars, and other analogous UV-bright enviroments.
The neglect of line forces in such 
disks in the past is probably due to a conceptual
bias toward point stars and radial radiation streams, which would not 
encounter velocity gradients and would not yield important line forces.
However, finite stellar
disks produce nonradial radiation streams that do encounter supersonic velocity
gradients even when there is no radial velocity at all.
The zeroth-order effects on disk structure may be negligible in high-density
regions unless $\alpha \ltwig 0.3$, but will likely become important
near the disk boundaries, where the densities are lower and closer to windlike.
This could yield important complications for disk winds, which might emanate
from disk material that requires a lower specific angular momentum to be in
orbit once radiative levitation is included.
Also, there can be an impact on the farther reaches of the disk where the
density is thought to drop steeply.
Finally, there are also important implications about the difficulty of building
a disk slowly from initially low densities, since 
the gas might not be containable against
radiative driving unless the density is high enough to yield
substantial self-shadowing in the lines.
Consequently, if constant line parameters
are adopted, the radiative force has greater impact for lower density,
lower $\alpha$, and higher $\Gamma$.

A central conclusion is that, notwithstanding
uncertainties in the appropriate parameters, the similarities in the
driving efficiency in a wind and in a disk strongly suggest that, in the absence of
additional disk confinement such as from magnetic fields,
it should not be possible to support a {\it low}-density Keplerian disk 
at the equator {\it and} a line-driven wind
over the poles
at the same time.
High-density disks, on the other hand, can avoid being
blown off like the polar winds only by virtue of
their reduction of line driving by self-shadowing.
As speculated by Marlborough (1997), non-steady mass input might
occur for Be disks in a process similar to miniature LBV eruptions.
Perhaps this could help increase the disk densities during their
vulnerable formation epoch.

Furethermore,
the potentially unstable role of self-shadowing in the presence of nonlinear
disk dynamics
suggests that radiative stripping of the disk might be able to
occur more catastrophically than 
the typical steady disk-wind model.
If so, it could contribute to
the commonly seen disappearances of disks,
although no such mechanism is offered here because of its
inherently nonlinear nature.
This is highly speculative, but the
possibility that nonsteady mass input to and output from a disk may
involve line driving justifies further studies into the role of radiative forces
in circumstellar disks.

In addition,
higher-order forces, such as the effects of gas pressure, quadrupole
gravity terms, and viscosity, should receive strong competition from
radial radiative forces, and also from radiative torques for noncircular
orbital perturbations.
Furthermore, the combined effect of radiative torque
and radial radiative force on a locally
aligned family of slightly elliptical orbits of fixed
eccentricity is to 
induce precession, and for suitable parameters the timescales can be
comparable to observations.
However, it is significant that the two effects are typically of similar
magnitude but opposite sign, and therefore compete.
The simple models calculated here predict that the retrograde precession
due to the force gradient should be stronger than the prograde precession
due to the torque effect, contrary to observations which favor prograde precession
of nonlinearly developed one-arm modes.
The degree of cancellation between
these effects might also contribute to the apparent stochasticity
of observed precession rates.

It must be cautioned that
all precession rates inferred here
are entirely schematic, since no global modes are included.
Eigenvalue calculations are needed to ascertain the impact on the
formation and evolution of small-amplitude one-arm modes, and
numerical calculations are needed to follow these into the nonlinear regime.
However, the results here suggest that dynamically sensitive line forces can play
an important role in such calculations, and should be included.

The quantitative reliability of all these conclusions is limited by the 
current uncertainties in the line parameters
in a Be-star disk, and by the absence of nonlinear hydrodynamic simulations
that include the radiative force.
The main
purpose of this paper has been simply to
categorize how these forces might be expected to alter disk
morphology, using an analysis of nearly circular orbits.
The extremely variable behavior of Be-star disks indicates that additional
physics will be required to understand them, but it seems likely that 
radiative forces will play a role in any complete description, especially
for the hotter and brighter stars.
The possibility also exists to extend these results to other UV-bright
quasi-Keplerian disks, such as accretion disks.
Recent advances in 
hydrodynamic modeling offer promise for future developments, as long as
the potential role of line forces is properly considered.

The authors would like to acknowledge helpful and interesting discussions with
Jon Bjorkman and Karen Bjorkman.
This work was supported in part by NASA grants 
NAG5-3530 and NAG5-4065.

\section{Appendix}

Here we derive the results, used in the text, 
that a torque proportional to $v_r$ yields
prograde precession over $N_\phi$ orbits, with
\beq
\label{nphiform}
N_\phi \ = \ {\Omega v_r \over g_\phi} \ = \ {1 \over 2} {g_{cor} \over
g_{\phi}} \ , 
\eeq
whereas 
a radial acceleration that is
proportional to $r^{-m_{rad}}$ over an orbit 
yields retrograde precession over $N_r$ orbits,
with 
\beq
\label{nrformapp}
N_r \ = \  {2 \over (2-m_{rad})} {r \Omega^2 
\over g_r} \ = \ 
{2 \over (2-m_{rad})} {g_{grav} \over g_r} \ .
\eeq
As in the text, $\Omega \ = \ \sqrt{g_{grav}/r}$ is the orbital frequency,
and $g_{cor} = 2 \Omega v_r$ is the magnitude of the coriolis force.
The sign convention is that positive $N$ implies prograde precession,
which holds for $N_\phi$ at suitably large radii, and for $N_r$ if
$m_{rad} < 2$.


The key expression for the azimuthal angle traced out during one 
passage from the inner turning point $r_p$ to the outer turning point $r_a$
can be written
\beq
\label{keyeq}
\Delta \phi = \int_{r_p}^{r_a} dr \ { \ang(r) \over r^2 v_r} \ ,
\eeq
where $\ang(r)$ is the specific angular momentum of the gas parcel.
This equation derives directly from $d\phi/dr = v_\phi/(r v_r)$ and
$v_\phi = \ang/r$.
The extent of the prograde or retrograde precession in each
orbit depends on how much
$\Delta \phi$ is greater or less than $\pi$.
The limits of the integration are the turning points where $v_r = 0$, which is
where the denominator is singular.
This singularity requires that the expression be evaluated carefully.

The radial speed $v_r$ is inferred from the energy equation, including
the work done by gravity, by the 
radial radiative force, and by the radiative torque.
To accomodate the effects of precession due to
a quadrupole moment, we split gravity into its usual point-mass
form 
$g_{grav}   =  -GM / r^2 $
plus a quadrupole correction $-f_q$ which scales as $r^{-4}$, where
the sign of $f_q$ is explicitly positive.
The radial radiative acceleration is denoted $f_r$, which varies
both as a function of $r_p$ and $r$, where $r_p$ is the periastron distance of
the orbital path in question, and $r$ specifies the varying location {\it along}
that path.
Thus variations from orbit to orbit are kept track of {\it separately}
from the variation over a given orbit.
In the same spirit, we further define two functions that vary over an orbit and are
necessary to accomodate the torque effect, which are
\beq
\label{fldef}
f_l(r) \ = \ {\ang_p r \over r_p^2}  {g_\phi \over v_r} 
\eeq
and
\beq
\label{fphidef}
f_\phi(r) \ = \ {\ang(r) \over r} {g_\phi \over v_r} \ ,
\eeq
where the subscript $p$ applies to quantities that are held fixed over the
orbit, say at their periastron values.
These functions will emerge shortly as key integrands for the angular momentum
and the kinetic energy, respectively.

The torque equation then allows us to specify $\ang(r)$ over an orbit as
\beq
\label{angeq}
\ang(r) \ = \ \ang_p \ + \ {r_p^2 \over \ang_p} \int_{r_p}^r dr' \ f_l(r') \ ,
\eeq
and the work integral then gives
\beq
{1 \over 2} \left ( v_r^2 \ + \ v_\phi^2 \right ) \ = \
{\ang_p^2 \over 2 r_p^2} \ - \ {\ang(r)^2 \over 2 r^2} \ + \
\int_{r_p}^r dr' \ \left [ g_{grav}(r') \ - \  f_q(r') \ + \ f_\phi(r')  \ + \ 
f_r(r') \right ] \ .
\eeq
The approach now is to substitute using equation (\ref{angeq}) and then
write an expression for the product $r v_r$, which can be written as
a quadratic in $r$ if all terms of order the (small) force corrections
$f_l$, $f_\phi$, $f_r$ and $f_q$ are Taylor expanded to second order in the small
quantity $r-r_p$.
Since all four $f$ functions appear
only as integrands, a second-order expansion merely requires
specifying their first derivatives $f'$, and we obtain
\beq
r v_r \ = \ \sqrt{r-r_p} \sqrt{r_a - r} \sqrt{
{2 GM \over r_p} \ - \ {l_p^2 \over r_p^2} \ + \ 
r_p^2 ( f'_l - f'_\phi ) \ - \ 4 r_p (f_\phi + f_r - f_q) \ - \ 
r_p^2 (f'_r - f'_q)} \ ,
\eeq
where the apastron distance $r_a$ is to order $f'$
\beq
r_a \ = \ r_p \left [
{l_p^2 \ - \ 2 r_p^3 f_l \ + \ r_p^4 f'_l \ - \ 2 r_p^3 ( f_\phi + f_r - f_q) \ - \ 
r_p^4 (f'_\phi + f'_r - f'_q) \over
2GM r_p \ - \ l_p^2 \ + \ r_p^4 f'_l \ - \ 4 r_p^3 (f_\phi + f_r - f_q) \ - \ 
r_p^4 (f'_\phi + f'_r - f'_q) } \right ] \ .
\eeq 

All this allows us to write equation (\ref{keyeq}) to lowest nontrivial order
in the form
\beq
\label{dphi}
\Delta \phi \ \cong \ 
{l_p \over \sqrt{l_p^2 - 2 r_p^3 f_l + r_p^4 f'_l - 2 r_p^3 (f_\phi+f_r-f_q) -
r_p^4(f'_\phi + f'_r - f'_q)}} \int_{r_p}^{r_a} {dr \over r} \
{\sqrt{r_a - r_p} \over \sqrt{r-r_p} \sqrt{r_a-r}} \ ,
\eeq
where the elementary integral on the right simply yields $\pi$.
As a check, note that if all $f$ functions become negligibly small we recover 
the purely Keplerian result of $\pi$ radians between turning points.
If we further define $N$ to be the number of orbits per precession cycle, then
$\Delta \phi \ = \ \pi (1-1/N)$, where $N > 0$ implies prograde precession.
Expanding equation (\ref{dphi}) then yields to lowest order
\beq
N \ \cong \ r_p \Omega_p^2 \left [
f_l \ + \ f_\phi \ + \ f_r \ - \ f_q \ + \ 
{r_p \over 2} ( f'_\phi - f'_l + f'_r - f'_q) \right ]^{-1} \ ,
\eeq
where $\Omega_p = l_p/r_p^2$ is the orbital angular velocity.

Thus all that remains is to evaluate the $f$ functions and their derivatives.
In present context, we have $f_\phi \ = \ f_l r_p^2/r^2$, so
$f'_\phi  -  f'_l \ = \ -2 f_l/r_p$ and $f_l = f_\phi$ when evaluated
at $r_p$, so we obtain simply
\beq
N \ = \ {2 r_p \Omega_p^2 \over \left [
2 (f_l  +  f_r - f_q) \ + \ r_p (f'_r - f'_q) \right ]} \ .
\eeq
The values of $f_l$ and $f_r$ at $r_p$ can be obtained by
substituting equations (\ref{gtograv})
and (\ref{torque}) into
equations (\ref{fldef}) and (\ref{fphidef})
respectively.
If we also define the exponent
$m_{rad}$ as the negative logarithmic derivative of $f_r$ with respect to $r$ at
$r_p$, such that
\beq
m_{rad} \ = \ - {r_p f'_r \over f_r} \ ,
\eeq
then we have $2f_r + r_p f'_r \ = \ (2-m_{rad}) f_r$.
For the gravitational correction, $f_q$ arises from knowledge of the
quadrupole moment of the stellar mass, and $f'_q \ = \ -4 f_q/r_p$.
Replacing periastron quantities with local values, we finally find
\beq
\label{nprecess}
N \ = \ {2 r \Omega^2 \over \left [ 2 f_l + (2-m_{rad})f_r + 2 f_q \right ]}  
\eeq
as the local expression for the precession period of embedded orbits with constant
eccentricity and parallel lines of nodes.

The denominator in equation (\ref{nprecess})
clearly specifies the separate precessional influences of
radiative torque, radial radiative forces, and quadrupole effects.
Since a positive $N$ implies prograde precession,
the torque effect yields prograde precession whenever the coefficient of
$v_r$ in $g_\phi$ (see eqs. [\ref{torque}] and [\ref{pphi}]) is positive.
The influence of radial gradients only yields prograde precession in limited
cases where
$m_{rad} < 2$, i.e., when the radiative force falls off more slowly than
gravity as a gas parcel moves over an eccentric orbit.
By contrast, the quadrupole term always yields prograde precession.

The fact that observed precession periods correspond to $\sim 10^3$ 
orbital periods at several stellar radii into the disk, coupled with
equation (\ref{nprecess}), implies that the precession may be caused either
by a non-inverse-square force with a magnitude of about $10^{-3}$ that
of gravity, or else by an azimuthal force of this same magnitude 
multiplied by the
ratio $v_r/v_\phi$.
Either way, the force perturbation must be quite small.
This supports the likelihood that radiative forces over the bulk of the
disk are indeed small, but does not preclude their potential to be large
in regions of low density.

 

\blankline
\noindent{\bf References}
\blankline

{


\noindent Babel, J. 1996, A\&A, 309, 867

\noindent Batchelor, G. K. 1967, An Introduction to Fluid Dynamics (Cambridge:
Cambridge Univ. Press), 598

\noindent Berio, P., Stee, Ph., Vakili, F., Mourard, D., Bonneau, D.,
Chesneau, O., Thureau, D., \& Hirata, R. 1999, A\&A, 345, 203



\noindent Castor, J. I., Abbott, D. C., \& Klein, R. I. 1975, ApJ, 195, 157 (CAK)


\noindent Chen, H, \& Marlborough, J. M. 1994, ApJ, 427, 1005


\noindent Cranmer, S. R. \& Owocki, S. P. 1996, ApJ, 462, 469

\noindent Cranmer, S. R. \& Owocki, S. P. 1995, ApJ,  440, 308

\noindent Cranmer, S. R., Smith, M. A., \& Robinson, R. D. 2000, ApJ, 537, 433

\noindent Dachs, J. 1987, in 
Physics of Be stars, Proceedings of the Ninety-second IAU Colloquium,
ed. A. Slettebak and T. Snow
(Cambridge: Cambridge Univ. Press), p. 149

\noindent Drew, J. E. \& Proga, D. 2000, NewAR, 44, 21



\noindent Feldmeier, A., Shlosman, I., \& Vitello, P. 1999, ApJ, 526, 357

\noindent Friend, D. B., \& Abbott, D. C. 1986, ApJ, 311, 701

\noindent Gayley, K. G. 1995, ApJ, 454, 410

\noindent Gayley, K. G. 2000, ApJ, 529, 1019

\noindent Gayley, K. G., \& Owocki, S. P. 2000, ApJ, 537, 461



\noindent Grinin, A. 1978, Sov. Astr., 14, 113

\noindent Hanuschik, R. W. 1994, Ap\&SS, 216, 99

\noindent Hanuschik, R. W., Hummel, W., Dietle, O., \& Sutorious, E. 1995, A\&A, 300, 163


\noindent Hubert, A. M. 1994, in IAU Symp. 162, Pulsation, Rotation and Mass Loss in 
Early-Type Stars, ed. L. A. Balona, H. F. Henrichs, 
\& J. M. Le
Contel (Dordrecht: Reidel), 341 

\noindent Hummel, W. \& Hanuschik, R. W. 1997, A\&A, 320, 852

\noindent Hummel, W. \& Vrancken, M. 2000, A\&A, 359, 1075



Utrecht.

\noindent Kudritzki, R. P., Pauldrach, A., Puls, J., \& Abbott, D. C. 1989,
 A\&A, 219, 205

Herrero, A. 1999,  A\&A, submitted


\noindent Lee, E. \& Goodman, J. 1999, MNRAS, 308, 984


\noindent Marlborough, J. M. 1997, A\&A, 317, L17

\noindent Mennickent, R. E., Sterken, C., \& Vogt, N. 1997,
A\&A, 326, 1167


\noindent Okazaki, A. T. 1991, PASJ, 43, 75

\noindent Okazaki, A. T. 1996, PASJ, 48, 305

\noindent Okazaki, A. T. 1997, A\&A, 318, 548

\noindent Okazaki, A. T. 2000, The Be Phenomenon in Early-Type Stars, 
eds. M. A. Smith, H. F. Henrichs, and J. Fabregat



\noindent Owocki, S. P., Cranmer, S. R., Gayley, K. G. 1996, ApJ, 472, L115.


\noindent Papaloizou, J. C., Savonije, G. J., \& Henrichs, H. F. 1992,
A\&A 265, L45

\noindent Pauldrach, A.W.A., Puls, J. \& Kudritzki, R.-P. 1986, A\&A, 164, 86






\noindent Proga, D., Stone, J., \& Kallman, T. R. 2000, ApJ, 543, 686



\noindent Puls, J., Springmann, U., \& Lennon, M. 2000, A\&AS, 141, 23

\noindent Quirrenbach, A., Bjorkman, K. S., Bjorkman, J. E., Hummel, C. A.,
Buscher, D. F., Armstrong, J. T., Mozurkewich,
D., Elias, N. M. II, \& Babler, B. L. 1997, ApJ, 479, 477
 

\noindent Rivinius, Th., Baade, D., Stefl, S., Stahl, O., Wolf, B.,
\& Kaufer, A. 1998, A\&A, 333, 125

\noindent Robinson, R. D. \& Smith, M. A. 2000, 540, 474


\noindent Smith, M. A., Grady, C. A., Peters, G. P., \& Feigelson, E. D. 1993,
ApJ, 409, L49


\noindent Telting, J. H., Waters, L. B. F. M., Roche, P., Boogert, A. C. A.,
Clark, J. S., de Martino, D., Persi, P. 1998, MNRAS, 296, 785

\noindent Telting, J. H., Heemskerk, M. H. M., Henrichs, H. F., \& Savonije, G. J.
1994, A\&A, 288, 558



\noindent Vakili, F., Mourard, D., Stee, Ph., Bonneau, D., Berio, P., Chesneau,
O., Thureau, N., Morand, F., Labeyrie, A., Tallon-Bosc, I. 1998, A\&A, 335, 261

\noindent Waters, L. B. F. M. 1986, A\&A, 162, 121

\noindent Waters, L. B. F. M. \& Marlborough, J. M. 1994, in IAU Coll. 162, 399

\noindent Wood, K., Bjorkman, K. S., \& Bjorkman, J. E. 1997, ApJ, 477, 926


}

\end{document}